\documentclass[aps,prl,reprint]{revtex4-1}

\usepackage[T1]{fontenc}
\usepackage[utf8]{inputenc}
\usepackage{amsmath}
\usepackage{mathtools}
\usepackage{graphicx}
\usepackage{hyperref}
\usepackage{datetime}
\usepackage{natbib}
\usepackage[font=small,labelfont=bf,tableposition=top]{caption}

\DeclareMathOperator{\EllipticK}{K}
\DeclareMathOperator{\EllipticE}{E}
\DeclareMathOperator{\EllipticPi}{\Pi_{1}}

\DeclareCaptionLabelFormat{andtable}{#1~#2  \&  \tablename~\thetable}

\begin{document}

\title{Large-N ground state of the Lieb--Liniger model and Yang--Mills theory on a two-sphere}
\author{Daniel Flassig}
\email{daniel.flassig@physik.lmu.de}
\author{Andre Franca}
\email{andre.franca@physik.lmu.de}
\author{Alexander Pritzel}
\affiliation{Arnold Sommerfeld Center, Ludwig-Maximilians-University, Theresienstr. 37, 80333 Munich, Germany}
\date{\formatdate{5}{8}{2015}}
\preprint{LMU-ASC 55/15}


\begin{abstract}

We derive the large particle number limit of the Bethe equations for the ground state of the attractive one-dimensional Bose gas (Lieb--Liniger model) on a ring and solve it for arbitrary coupling. We show that the ground state of this system can be mapped to the large-$N$ saddle point of Euclidean Yang--Mills theory on a two-sphere with a $U(N)$ gauge group, and the phase transition that interpolates between the homogeneous and solitonic regime is dual to the Douglas--Kazakov confimenent-deconfinement phase transition.

\end{abstract}

\maketitle

\section{Introduction}

The Lieb--Liniger (LL) model is an interesting laboratory to study properties of strongly interacting quantum many body systems, both experimentally \cite{Strecker, Khaykovich, Paredes, Haller} and theoretically. The attractive version has been used extensively to study, e.g., quench dynamics \cite{Calabrese, Calabrese1, CalabreseKPZ, Caux1, Caux} and is known to undergo a phase transition  at large particle number \cite{Ueda}. Yet at the same time, the model is integrable and can be solved exactly using the Bethe ansatz \cite{Lieb}. 

In practice however, a closed form expression for the Bethe state has only been available in the weak and strong coupling limits. Therefore studies of the phase transition have mostly resorted to mean field methods or numerical diagonalization of the Hamiltonian (see, e.g., \cite{Sykes, Sakmann, Flassig, Goldstone}).

In this letter, we derive the continuum limit of the Bethe equations for the ground state of the attractive LL model and solve it for arbitrary coupling. We confirm the second order phase transition by considering the ground state energy \footnote{A finite temperature liquid--gas phase transition of first order has also been observed recently \cite{Herzog}. }.

Finally we observe that the ground state can be mapped exactly to the large-N saddle point of U(N) Yang--Mills theory on a two-sphere, where the phase transition manifests itself as the confinement-deconfinement phase transition of Douglas and Kazakov \cite{Douglas}, which is deeply connected to random matrix theory \cite{SchehrMatrix} and has diverse manifestations \cite{SchehrBrownian, Pronko}.
\begin{figure*}[t]

        \centering
       \includegraphics[width=0.19\linewidth]{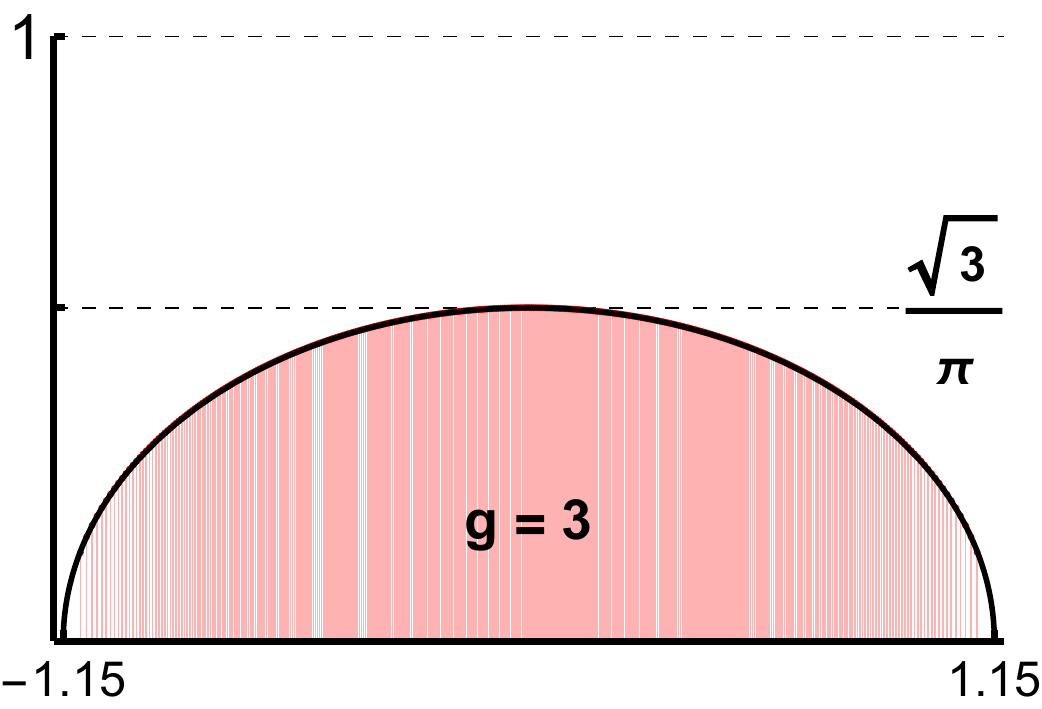}
       \includegraphics[width=0.19\linewidth]{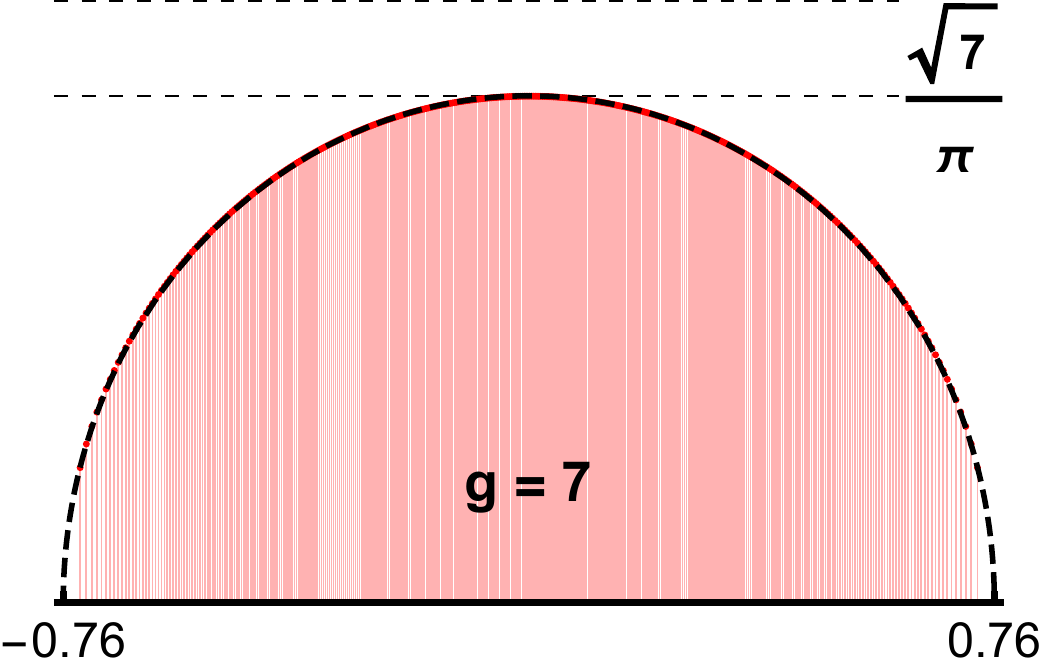}		
	   \includegraphics[width=0.19\linewidth]{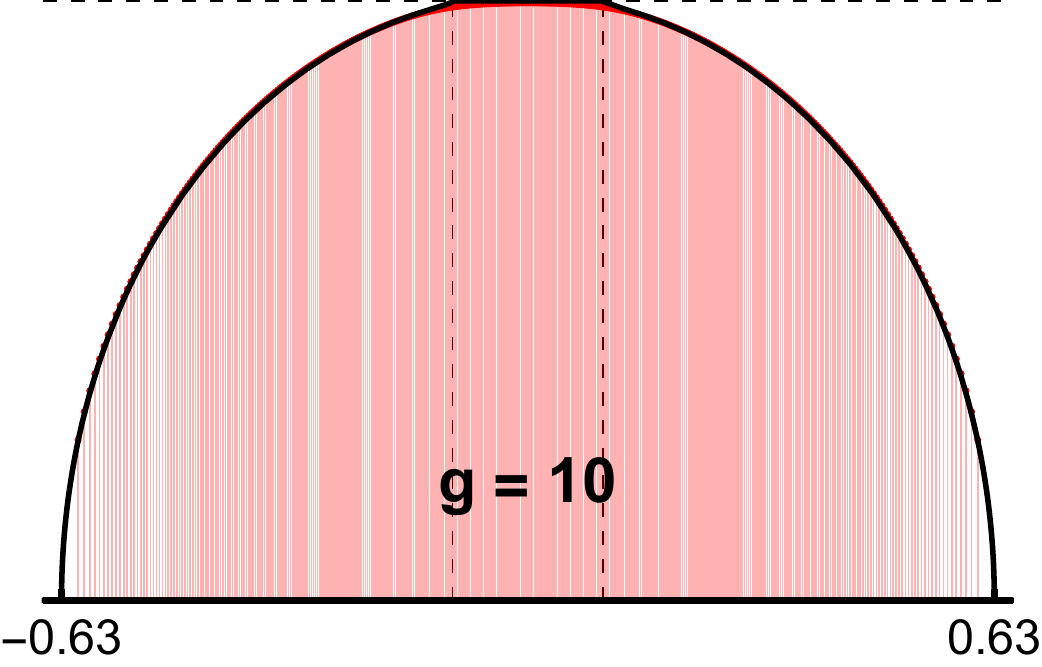}
	   \includegraphics[width=0.19\linewidth]{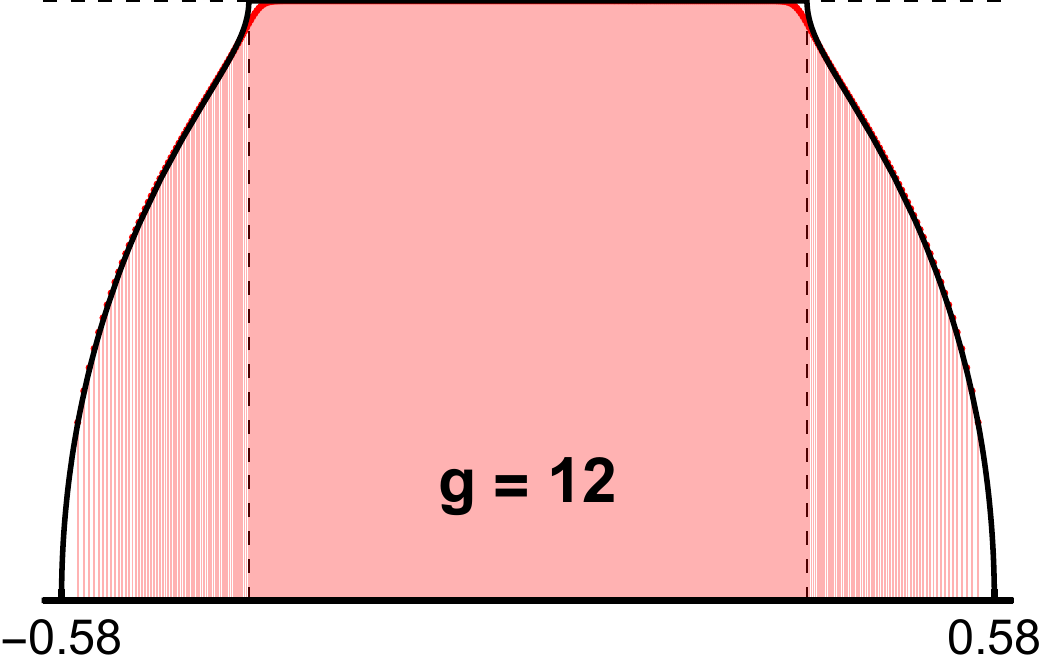}
	   \includegraphics[width=0.19\linewidth]{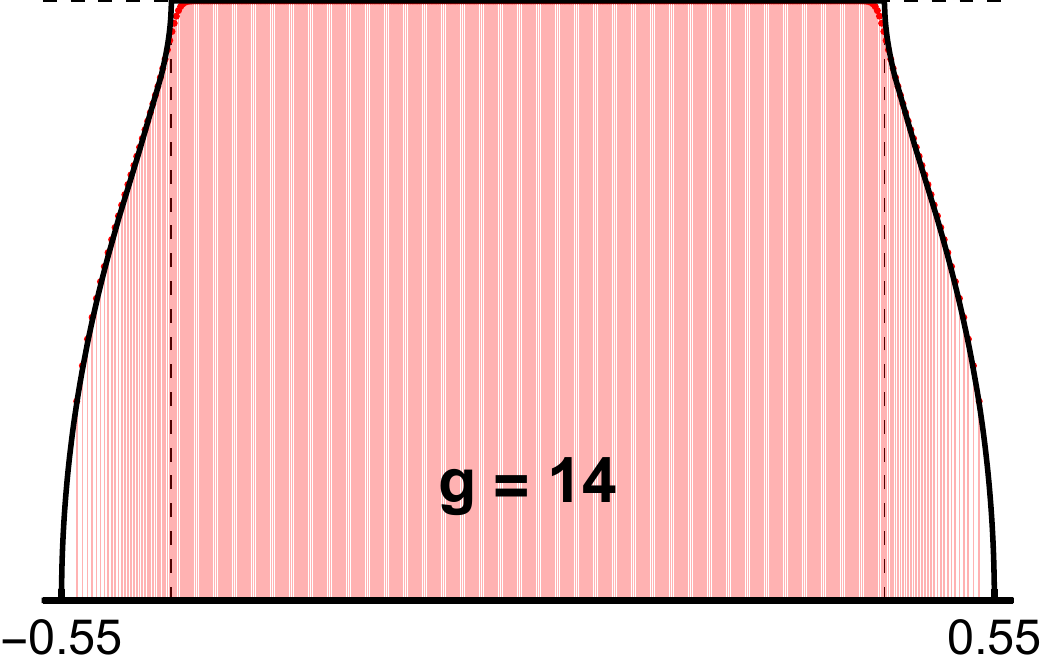}
	\caption{Continuum root distribution $\rho = 1/k'$ (black) and numerical data points at $N=400$ (red fill) }
	\label{fig:density}
\end{figure*}
\section{Bethe ansatz for the ground state}\label{sec:bethe-review}

We are interested in the Lieb--Liniger Hamiltonian \cite{Lieb} 
\begin{equation}
H=-\sum_{i\leq N}\frac{\partial^{2}}{\partial x_{i}^{2}}-c\sum_{i\neq j}\delta\left(x_{i}-x_{j}\right)\label{eq:H}
\end{equation}
for $N$ identical Bosons on the interval $[0, L)$ with periodic boundary conditions. We use the sign convention in which the coupling is attractive for $c>0$. 

The eigenstates of this Hamiltonian can be computed exactly using the Bethe ansatz \cite{Lieb, Korepin}. They are fully characterized by a set of complex numbers $k_j$, roots of the Bethe equations
\begin{equation}
	\label{eq:beq}
	\text{e}^{ik_i L}=\prod_{j\neq i}\frac{k_i-k_j-ic}{k_i-k_j+ic}.
\end{equation}		
The energy and momentum eigenvalues of the Bethe states are $E = {\sum}_i \: k_i^2$, $P = {\sum}_i \: k_i$.

In the repulsive regime, it was shown \cite{Yang}
that the Bethe states with real roots form a complete set of the $N$-particle
Hilbert space. For attractive $c>0$, there exist bound states: subsets
of roots with identical real but differing imaginary
part. The ground state of the system is then characterized by a bound state of zero real momentum, with all the $k_i$ purely imaginary \cite{McGuire}.

Replacing $k_{j}\rightarrow -ik_{j}$, we then arrive at the form of the Bethe equations that we will deal with:
\begin{equation}
	\label{eq:beqgs}
	k_i L=\sum_{j\neq i}\log\frac{k_i-k_j+c}{k_i-k_j-c}
\end{equation}
Their real solution characterizes the ground state of the system.

In the thermodynamic limit, where $N$ and $L$ are taken to infinity with fixed density $N/L$, the Bethe roots form an exact string with exponentially small deviations: $k_{j} \approx c(j - (N+1)/2)$ \citep{McGuire}. This limit, however, is inherently a strong coupling limit, which can be seen by noticing that by using dimensionless roots in the Bethe equations, the coupling becomes $cL$.

At weak coupling for $c\rightarrow 0$ ($N$ fixed), the distance between adjacents roots is much larger than $c$ and the roots are distributed according to Wigner's semi-circle law  \cite{Oelkers}. 

In both limits, the roots obey the following inequality that is a direct consequence of \eqref{eq:beqgs}, but will have to be imposed in the continuum limit. 
\begin{equation}
	\label{eq:constraint}
	|k_i - k_j| > c 
\end{equation} 

To see why this inequality holds, imagine changing the coupling adiabatically. Assume that at some point $k_{i+1}-k_i = c$ for a pair of roots (if there are several such pairs, focus on the one with lowest index). The sum in the Bethe equation (\ref{eq:beqgs}) for $k_i$ has just one diverging contribution at this point and can no longer be satisfied. 

\section{Large-$N$ limit}
From here on, we will consider the large $N$ limit of the LL model at finite effective coupling $g$
\begin{equation}
	N \rightarrow \infty \quad \text{while} \quad g = c L N = \mathrm{const}
\end{equation}
which is the correct limit to observe the phase transition. We also set $L=1$ for convenience. It turns out that the root distribution converges to a continuous function in this limit if we define 
\begin{equation}
	\label{eq:continuum-k}
	k_i \equiv g \, k(i/N)
\end{equation}

The sum on the rhs of the Bethe equation can be split in a near contribution from $|j-i|<\epsilon N$ and the rest, for some $\epsilon > 0$. A closer analysis reveals, that for $k' > 1$, the near contribution vanishes in the double limit $\lim_{\epsilon \rightarrow 0} \lim_{N \rightarrow \infty}$ (see appendix). For the rest of the sum, the following Taylor expansion is valid
\begin{equation}
\label{eq:c1}
\lim_{N \to \infty} \log \frac{k_i - k_j + \frac{g}{N} }{k_i - k_j - \frac{g}{N} } =  \frac{ 2 g }{N} \frac{1}{k_i - k_j} .
\end{equation}

In the continuum limit, the Bethe equation thus becomes an integral equation 
\begin{equation}\label{eq:bethe-continuum}
	g k = 2\: \mathcal{P}\!\!\int_{- k_{\mathrm{min}}}^{k_{\mathrm{max}}} \frac{\rho(u)}{k - u} \mathop{d u}
\end{equation}
where $\rho(k) \equiv 1/k'$ is the density of roots and the principal value symbol is a remnant of the $\epsilon$ excision. The bounds must be chosen such that $\int_{-k_{\mathrm{min}}}^{k_{\mathrm{max}}} \rho(u) \mathop{du} = 1$, and the density must satisfy the constraint (\ref{eq:constraint}), which in the continuum limit becomes
\begin{equation}
\label{eq:rhoconstraint}
\rho(k)\leq1
\end{equation}

The solution of the integral equation (\ref{eq:bethe-continuum}) is Wigner's semi-circle law
\begin{equation}
\label{eq:semicircle}
\rho(k)=\frac{1}{\pi}\sqrt{g-\frac{g^{2}k^{2}}{4}}
\end{equation} 

As long as $g < \pi^2$, the constraint (\ref{eq:rhoconstraint}) is satisfied and (\ref{eq:semicircle}) is the correct ground state root distribution. For $g>\pi^2$, however, this distribution violates the constraint. 

Studying the large $N$ limit of the exact string solution reveals, that the continuum root distribution may saturate the constraint $\rho(k) = 1$ on an interval $k \in [-b,b]$. We therefore make the following ansatz in the solitonic regime
\begin{equation}
\rho(k)=\begin{cases}
1 & k \in [-b,b]\\
\tilde{\rho}(k) &  k \in [-a,-b)\cup(b,a]
\end{cases}
\end{equation}

Inserting into (\ref{eq:bethe-continuum}) then produces an integral equation for $\tilde{\rho}$, the solution of which is \cite{Pipkin, Douglas}
\begin{equation}
\label{eq:sol-soliton}
\tilde{\rho}(k)=\frac{2}{\pi a |k|} \sqrt{(a^2 - k^2)(k^2 - b^2)} \; \EllipticPi \Big(\frac{b^{2}}{k^{2}},\frac{b^2}{a^2} \Big)
\end{equation}
and the parameters $a$ and $b$ are determined from the following conditions:
\begin{equation}
\label{eq:params}
\begin{gathered}
4 \EllipticK(x) (2 \EllipticE(x) - (1-x) \EllipticK(x))=g\\
a g=4\EllipticK(x) \quad \text{and} \quad x=b^2/a^2
\end{gathered}
\end{equation}
where $\EllipticE(x)$ and $\EllipticK(x)$ are the elliptic functions of the first and second kind, and $\EllipticPi(x,y)$ is the elliptic function of the third kind \cite{WolframFunctions}, defined as \footnote{Note that conventions differ on the square of the second argument.}
\begin{equation}
\EllipticPi(x,y)=\int_{0}^{1} \frac{1}{(1 - x u^2) \sqrt{1 - y u^2} } \frac{\mathop{du}}{\sqrt{1 - u^2}}.
\end{equation}
Note that for $g \rightarrow \pi^2$, we have $b\rightarrow 0$ and $\tilde {\rho} (\pi^2) $ becomes a semi-circle and thus the root distribution changes continuously at the phase transition.

In figure \ref{fig:density} we show the continuum limit root distribution for several values of the effective coupling. The numerical results for $N=400$, obtained directly from (\ref{eq:beqgs}), are superimposed on the graphs and match very well.

\section{Ground state energy and phase transition\label{sec:phase-transition}}

In the large-$N$ limit, the energy per particle becomes
\begin{equation}
\epsilon = - \frac{1}{N} {\sum}_i \: k_i^2 = -g^2 \int k^2 \rho( k) \mathop{d  k}
\end{equation}

For the weak coupling solution (\ref{eq:semicircle}) this expression is simple to evaluate. On the strong coupling side (\ref{eq:sol-soliton}), the integral representation of $\EllipticPi$ and contour integration can be used to calculate the energy. After simplifying with (\ref{eq:params}), we get
\begin{equation}
\label{eq:sol-energy}
-\epsilon = 
	\begin{cases}
		g & \text{for} \quad g \le \pi^2 \\
		\frac{1}{48} g^2 \Big( 8 (a^2 + b^2) + g (a^2 - b^2)^2 \Big) & \text{for} \quad g > \pi^2
	\end{cases}
\end{equation}

By inverting (\ref{eq:params}) and expanding $a$, $b$ and finally $\epsilon$ as a power series in the effective coupling $g$ around $\pi^2$, we find
\begin{equation}
- \epsilon = g  +  \frac{2}{\pi^2} (g-\pi^2)^2  + \mathcal{O}\big((g-\pi^2)^3\big) \text{, for} \quad g > \pi^2
\nonumber
\end{equation}
We observe that $\epsilon ( g )$ and $\epsilon ' (g)$ are continuous at $\pi^2$, whereas the second derivative is discontinuous, confirming that it is indeed a second order phase transition. 

As a non-trivial check, we can compare (\ref{eq:sol-energy}) with the expression obtained in \cite{Ueda} using mean field theory, since in the large-$N$ limit we expect mean field to produce the correct ground state energy \footnote{Right at the phase transition point, mean field theory (and the Bogoliubov approximation) can never give the correct result. At any finite distance from the phase transition however, $N$ can be chosen sufficiently large to make the approximation arbitrarily good.}. In our conventions, the mean-field ground state energy is, for $g>\pi^2$
\begin{equation}
\label{eq:mf-energy}
-\epsilon_\text{mf} = \frac{4}{3}\frac{\EllipticK(m)^2}{\EllipticE(m)}\big( (2-m) \EllipticE(m) + (1-m) \EllipticK(m) \big)
\end{equation}
with $m$ determined from
\begin{equation}
4 \EllipticE(m) \EllipticK(m) = g
\end{equation}
and although we have not succeeded in proving the equivalence of the two expressions (\ref{eq:sol-energy}) and (\ref{eq:mf-energy}) by algebraic means, we have checked their numerical equality at various values of $g$ to several dozen digits of precision.
\begin{figure}[t]
	\includegraphics[width=0.90\linewidth]{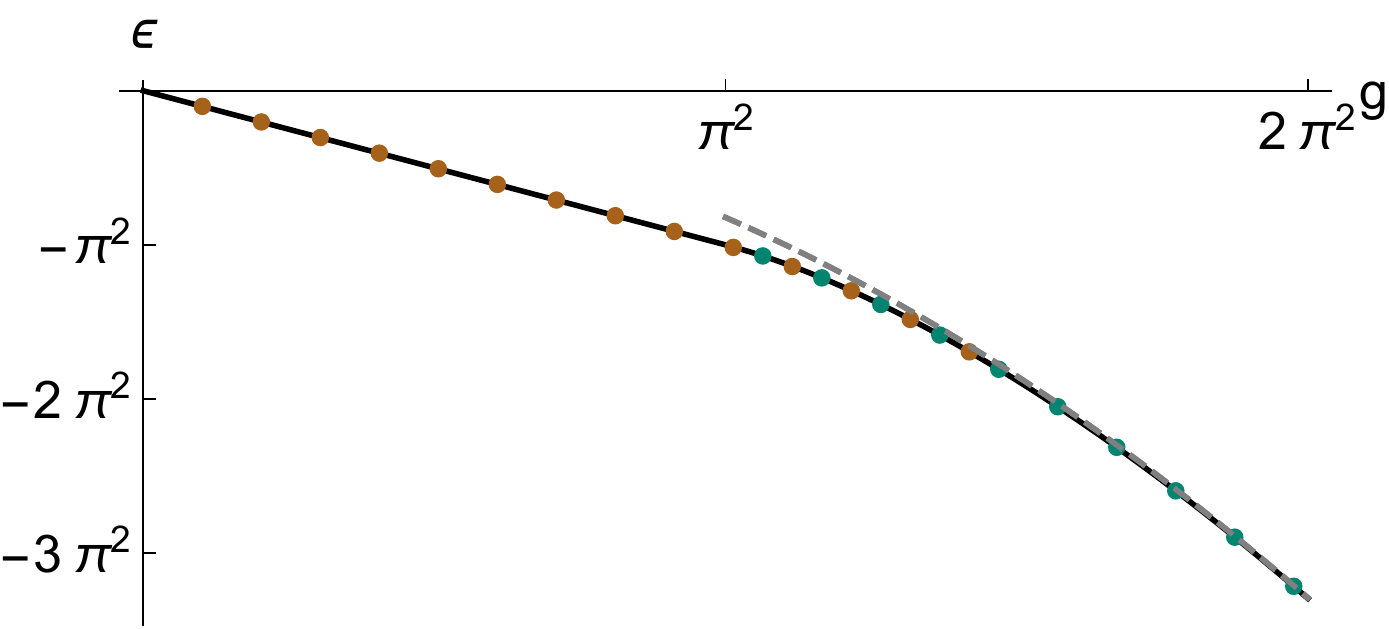}
    \caption{Ground state energy per particle. Numerical results for 400 particles are shown in brown. In green is the mean-field result in the strong coupling phase. The dashed line shows the thermodynamic limit \cite{McGuire}.}
\end{figure}
\section{Equivalence to Yang--Mills theory on a two-sphere\label{sec:yang-mills}}

We will briefly recap the large $N$ limit of the $U(N)$ Yang--Mills partition function in two dimensions as derived in \cite{Douglas}. It will then be obvious how our ground state of the Lieb--Liniger model maps directly to this theory quantized on a sphere.

The partition function of pure Yang--Mills theory on a two dimensional manifold of genus $G$ and area $A$ can be expressed as a sum over representations $R$ of the gauge group \cite{Rusakov}
\begin{equation}
Z_{G}\left(A\right)=\sum_{R}(\dim R)^{2-2G} e^{-A \lambda^2 C_2(R)/2N}
\end{equation}
where $\lambda$ is the 't Hooft coupling.	

For the gauge group  $U(N)$, the sum over representations can be expressed as a sum over Young tableaux characterized
by a set of decreasing integers $\{n_{1},n_{2},...,n_{N}\}$, the components of the
highest weight.

In the 't Hooft large-$N$ limit the representations may be characterized by a continuous function $h$:
\begin{equation}
	N \, h(i/N) \equiv -n_i+i-N/2
\end{equation}
and the partition function becomes
\begin{align}\label{eq:ym-part-large-n}
	Z_{G=0}(A)=&\int Dh(x) \exp(-N^2 S_{\mathrm{eff}}[h]) 
	\nonumber \\
	S_{\mathrm{eff}}[h]=-&\int_{0}^{1}\int_{0}^{1} \log |h(x)-h(y)| \,dx\,dy+
	\nonumber \\		
	&\frac{A \lambda^2}{2}\int_{0}^{1} h(x)^2\,dx-\frac{A \lambda^2}{24}
\end{align}

Since the $n_i$ are monotonic, it is clear that $h(x)$ obeys the inequality $h(x)-h(y) \geq x-y$, so $h'(x) \geq 1$.

The large-N saddle point approximation of (\ref{eq:ym-part-large-n}) yields an integral equation for the density $\rho(h) = dx/dh$
\begin{equation}
	\label{eq:ym-saddlepoint}
	A \lambda^2 h = 2\: \mathcal{P}\!\!\int \frac{\rho(s)}{h-s} \, ds
\end{equation}

Clearly this integral equation with constraint is identical to equation (\ref{eq:bethe-continuum}) that governs the Bethe root distribution in the ground state of the Lieb--Liniger model. The correspondence directly maps the density of Young tablaux boxes $h$ to the density of Bethe roots $k$ and the 't Hooft coupling $\lambda^2$ to the effective LL coupling $g$.

The phase transition at $g=\pi^2$ in the Lieb--Liniger model appears as the confinement/deconfinement phase transition in the gauge theory.

It is not yet clear, but an interesting open question, how physical observables of both systems can be related to each other.

Note that this is not the first time a correspondence between Bethe equations of an integrable one-dimensional system and Yang--Mills theory has been found \citep{Nekrasov2009}. In the known examples, the system was mapped into the moduli space of a supersymmetric gauge theory. And the Bethe roots played the role of the eigenvalues of the complex scalar in the vector multiplet. 

In our case, however, we map the Bethe roots to the components of  highest weight of the representation of $U(N)$ that dominates the saddle point of the partition function. 
\begin{figure}[t]
\begin{center}
\begin{minipage}[t]{.57\linewidth}
\vspace{0pt}
\centering
\includegraphics[width=\linewidth]{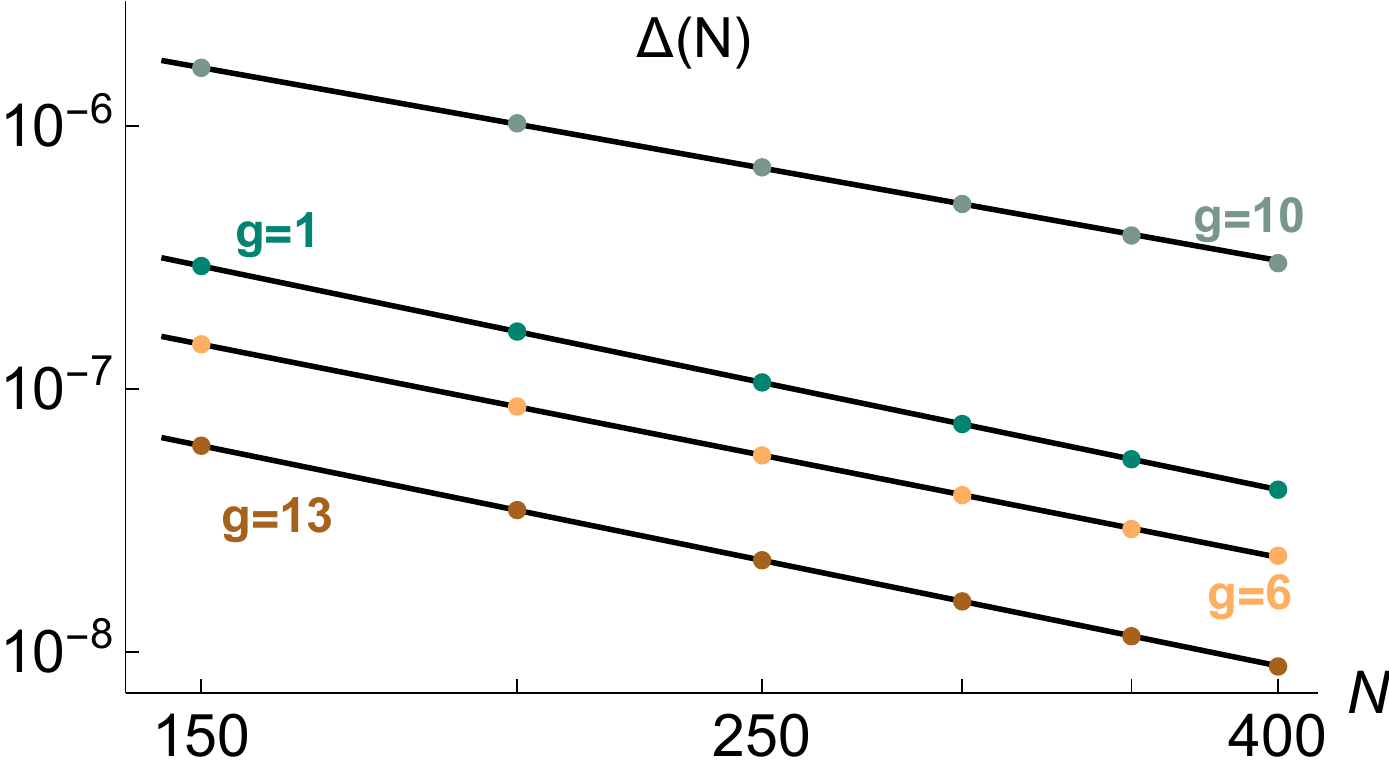}
\end{minipage} \,
\begin{minipage}[t]{.38\linewidth}
\vspace{0pt}
\centering
\scalebox{0.9}{\begin{tabular}{cccccc} \hline
	g & B & g & B & g & B			\tabularnewline \hline
	1 & 2.00 & 6 & 1.97 & 11 & 1.98	\tabularnewline
	2 & 2.00 & 7 & 1.95 & 12 & 1.94	\tabularnewline
	3 & 1.99 & 8 & 1.94 & 13 & 1.90	\tabularnewline
	4 & 1.99 & 9 & 1.91 & 14 & 1.87	\tabularnewline
	5 & 1.98 & 10 & 1.72 &  & 		\tabularnewline \hline
    \end{tabular} } 
\end{minipage}
\captionlistentry[table]{B(g)}
\captionsetup{labelformat=andtable}\label{tab:bofg}
\caption{\label{fig:bofg}Asymptotic behavior of $\Delta(N)$ with best fit parameters $B$ for different couplings.}
\end{center}
\end{figure}

\section{Numerical Checks\label{sec:numerics}}
We have performed numerical checks to validate our continuum results. To this end, we have solved the Bethe equations (\ref{eq:beqgs}) at various values of $N$ and $g$, using the Levenberg-Marquadt solver provided by Mathematica \footnote{As is usually done, we used a parametrization for the roots $k_j = j c + \delta_j$ in order to optimally exploit the floating point representation in the strong coupling phase.}. In order to probe the convergence of the finite $N$ root distribution to the analytic large $N$ expression, we compute the mean square deviation
\begin{equation}
\Delta\left(N,g\right)=\frac{1}{N}\sum_{i=1}^{N}\left(k_{i}\left(g\right)-g\bar{k}\left(\frac{i}{N},g\right)\right)^{2}
\end{equation}
where $\bar{k}(x,g)$ is defined by numerically integrating (\ref{eq:sol-soliton}, \ref{eq:semicircle}).

The results are displayed in Fig. \ref{fig:bofg}, where we show $\Delta\left(N,g\right)$ as a function of $N$ for different values of $g$. We observe that $\Delta\left(N,g\right)$ behaves like $A(g) N^{-B(g)}$. The best fit parameters $B(g)$ are shown in table \ref{tab:bofg}. Based on these numbers, we conjecture that $B = 2$ at large N, and we notice that subleading (in N) effects seem to be stronger around the phase transition $g=\pi^2$.

\section{Conclusions\label{sec:Conclusions}}
In this letter we have studied the attractive Lieb--Liniger model in the large-$N$ scaling limit. We have derived an integral equation (\ref{eq:bethe-continuum}) that provides the continuum form of the Bethe equations. Together with a bound on the root density, this has allowed us to calculate the limiting form of the Bethe root distribution (\ref{eq:semicircle}, \ref{eq:sol-soliton}). 

The phase transition from the homogenous weak coupling to the bright soliton phase manifests itself in a change in the functional form of the root distribution. The ground state energy (\ref{eq:sol-energy}) coincides with the mean-field result - involving an identity of elliptic integrals - and confirms the order of the phase transition (second order). 

The equivalence between the large-$N$ saddle point of $U(N)$ YM theory on a sphere with the scaling limit of the LL model seems like a promising avenue for future investigations, especially considering that various relations between (supersymmetric) YM theory and integrable systems have already been uncovered \cite{Nekrasov2009}. 

Another interesting direction is how to compute the lowest-lying excitations in the large-$N$ limit. The knowledge of the root density for the first excitations would allow us to probe the time evolution of observables at arbitrary couplings.

\begin{acknowledgements}
It is a pleasure to thank G. Dvali, N. Wintergerst and M. Panchenko for many valuable discussions regarding the physics of quantum phase transitions and the Lieb--Liniger model. We also want to thank C. Gomez and S. Hofmann for discussions and encouragement. 
The work of D.F. and A.P. was supported by the Humboldt Foundation. 
The work of A.F. was supported by the FCT through Grant No. SFRH/BD/77473/2011.
\end{acknowledgements}
\bibliographystyle{apsrev4-1}
\bibliography{references}
\appendix
\section{Appendix}

For some $\epsilon > 0$ the near contribution to the sum (\ref{eq:beqgs}) from indices $|j - i| \le \epsilon N$ can be written in the following form
\begin{equation*}
    \sum_{\mathclap{\delta = 1}}^{\epsilon N} \,\,
	\log \Bigg(1 + \frac{2}{N}\,\frac{k_{i-\delta} + k_{i+\delta} - 2 k_i}{( k_{i+\delta} - k_i + \frac{g}{N})( k_i - k_{i-\delta} - \frac{g}{N})} \Bigg)	
\end{equation*}

We switch to continuum variables (\ref{eq:continuum-k}), with $x=i/N$ and assume that $k'(x) > 1$ and $k''(x) \neq 0$. Defining $k'^2_\mathrm{min} = \min \, k'(y)^2$ for $y \in [x-\epsilon,x+\epsilon]$, the above sum can be bounded by 
\begin{equation*}
\sum_{\mathclap{\delta = 1}}^{\epsilon N} \,\, \log \Bigg(1 + \frac{2}{N} \frac{c |k''(x)|}{k'^2_\mathrm{min} - 1} \Bigg) \leq \epsilon \: \frac{c |k''(x)|}{k'^2_\mathrm{min} - 1}
\end{equation*}
for some $c > 1$. Now it is obvious that this near contribution vanishes in the limit $\lim_{\epsilon \rightarrow 0} \lim_{N \rightarrow \infty}$.

\end{document}